\documentclass{elsart}
\usepackage{graphicx}

\begin{document}

\def\dj{\leavevmode\setbox0=\hbox{d}\dimen0=\wd0
        \setbox0=\hbox{d}\advance\dimen0 by -\wd0
        \rlap{d}\kern\dimen0\hbox to \wd0{\hss\accent'26}}
\def\DJ{\leavevmode\setbox0=\hbox{D}\dimen0=\wd0
        \setbox0=\hbox{D}\advance\dimen0 by -\wd0
        \rlap{D}\kern\dimen0\hbox to \wd0{\raise -0.4ex\hbox{\accent'26}\hss}}

\begin{frontmatter}

\title{$\alpha$-decay of excited states in $^{11}$C and $^{11}$B}

\author[bham,irb]{N. Soi\'{c}},
\author[bham]{ M. Freer},
\author[bham]{ L. Donadille\thanksref{sac}},
\author[bham]{ N. M. Clarke},
\author[bham]{ P. J. Leask},
\author[surrey]{W. N. Catford},
\author[surrey]{ K. L. Jones\thanksref{gsi}},
\author[surrey]{ D. Mahboub},
\author[york]{ B. R. Fulton},
\author[york]{ B. J. Greenhalgh},
\author[york]{ D. L. Watson} \and
\author[anu]{D. C. Weisser}

\address[bham]{School of Physics and Astronomy, University of Birmingham, Edgbaston,
Birmingham B15 2TT, United Kingdom }
\address[irb]{Ru\dj er Bo\v{s}kovi\'{c} Institute, Bijeni\v{c}ka 54,
HR-10000 Zagreb, Croatia}
\address[surrey]{School of Electronics and Physical Sciences, University of Surrey, Guildford,
 Surrey, GU2 5XH, United Kingdom}
\address[york]{Department of Physics, University of York, Heslington, York, YO10 5DD, 
United Kingdom}
\address[anu]{Department of Nuclear Physics,  The Australian National University, Canberra ACT
 0200, Australia }

\thanks[sac]{Present address: CEA-Saclay, DAPNIA/SPhN, Bt. 703, Pice 162, F-91191 Gif sur
 Yvette Cedex, France}
\thanks[gsi]{Present address: GSI, Gesellschaft f\"{u}r Schwerionenforschung mbH,
 Planckstrasse 1, D-64291 Darmstadt, Germany}

\begin{abstract}
Studies of the $^{16}$O($^{9}$Be,$\alpha$$^{7}$Be)$^{14}$C and
$^{7}$Li($^{9}$Be,$\alpha$$^{7}$Li)$^{5}$He reactions at E$_{beam}$=70
MeV have been performed using resonant particle spectroscopy techniques. 
The $^{11}$C excited states decaying into $\alpha$+$^{7}$Be(gs) are
observed at 8.65, 9.85, 10.7 and 12.1 MeV as well as possible states at
12.6 and 13.4 MeV. This result is the first observation of $\alpha$-decay 
for excited states above 9 MeV. The $\alpha$+$^{7}$Li(gs) decay
of $^{11}$B excited states at 9.2, 10.3, 10.55, 11.2, (11.4), 11.8, 12.5, 
(13.0), 13.1, (14.0), 14.35, (17.4) and (18.6) MeV is observed. The decay
processes are used to indicate the possible three-centre 
2$\alpha$+$^{3}$He($^{3}$H) cluster structure of observed states. 
Two rotational bands corresponding to very deformed structures are 
suggested for the positive-parity states. Excitations of some observed 
T=1/2 resonances coincide with the energies of T=3/2 states which are 
the isobaric analogs of the lowest $^{11}$Be states.
Some of these states may have mixed isospin.
\end{abstract}


\end{frontmatter}

\section{Introduction}

It is well known that many light nuclei possess a prominent cluster structure
and that the $\alpha$-particle has an important impact on their structure. In 
recent years, special attention has focused on beryllium isotopes where well 
developed cluster structure was found in $^{8,9,10}$Be and tentative evidence 
for such behaviour was found in $^{11,12,14}$Be 
\cite[\ and references therein]{voe,freercrp}.  For example,
recent measurements have provided evidence for an $\alpha$+$^{6}$He cluster 
structure in $^{10}$Be \cite{so10be,freerpr,mil99,curtis,liendo} and for 
a $^{6}$He+$^{6}$He structure in $^{12}$Be \cite{freerpr}, and the unusual 
structural properties of $^{11}$Be have also attracted significant interest 
\cite{palit,bohlen,winfield,cappu}. Measurements of the helium-cluster breakup 
and neutron removal cross-sections suggest that neutron-rich Be isotopes 
possess a strong structural overlap with an $\alpha$+Xn+$\alpha$ configuration 
\cite{nia}. It appears that properties of beryllium nuclei may be well described
in terms of the sharing of the valence neutrons between the two $\alpha$-cores 
in a manner which is reminiscent of the covalent binding of atomic molecules. 
The presence of a 3$\alpha$ cluster structure in $^{12}$C provides an extension 
to this idea. The recent studies of the $\alpha$-decaying states in $^{13}$C
\cite{so13c} and $^{14}$C \cite{so14c} have found indications of molecular
structures in these nuclei and tentative evidence for the chain structure 
has been found in $^{13}$C \cite{mil02}.

It is interesting to investigate influence of $\alpha$-clustering on 
the structural properties of the neutron deficient nucleus $^{11}$C and also
on boron isotopes which are situated between the beryllium and carbon nuclei.
One particulary interesting issue is the existence of multi-centre 
structures in $^{11}$B, are they 
two-centre, as in Be isotopes, or three-centre as in C isotopes? Detailed 
knowledge of the structure of boron isotopes may help in understanding of the 
molecular nature of light nuclei and its evolution from two- to three-centre 
structures. Although $^{11}$C and particulary $^{11}$B nuclei have been studied 
extensively, the experimental evidence for cluster structures is rather scarce. 
It is worth mentioning that $\alpha$+$^{7}$Be and $\alpha$+$^{7}$Li reactions
and structure of $^{11}$C and $^{11}$B are also of considerable astrophysical 
interest: the $^{7}$Be($\alpha$,$\gamma$)$^{11}$C reaction is starting point 
of the hot pp chain and $^{7}$Li($\alpha$,$\gamma$)$^{11}$B is the main 
production process of $^{11}$B in the big-bang nucleosynthesis.
We present here results of the experimental studies which probe cluster 
structure of $^{11}$C and $^{11}$B via the $\alpha$-decay of their 
excited states. The $^{11}$C excited states have been studied using the 
$^{16}$O($^{9}$Be,$\alpha$$^{7}$Be)$^{14}$C reaction and study of the $^{11}$B
excited states has been performed using the 
$^{7}$Li($^{9}$Be,$\alpha$$^{7}$Li)$^{5}$He reaction. It is the two-nucleon
transfer processes onto the cluster nucleus $^{9}$Be which provides a possible 
mechanism by which the multi-centre cluster structures may be populated.

\section{Experimental Details}

The measurements were performed at the Australian National University's 
14UD tandem accelerator facility. A 70 MeV $^9$Be beam, of intensity 3 
enA, was incident on a 100 $\mu$gcm$^{-2}$ Li$_2$O$_3$ foil.

Reaction products formed by interaction of $^9$Be with the
target were detected in an array of four charged particle
telescopes. These telescopes contained three elements which
allowed the detection of a wide range of particle types, from
protons to $Z$=4 to 5 nuclei. The first elements were thin,
70$\mu$m, 5$\times$5 cm$^{2}$ silicon detectors segmented into
four smaller squares (quadrants). The second elements were
position-sensitive strip detectors with the same active area as
the quadrant detectors, but divided into 16 position-sensitive
strips. These strips were arranged so that the position axis gave
maximum resolution in the measurement of scattering angles.
Finally, 2.5 cm thick CsI detectors were used to stop highly
penetrating light particles. These detector telescopes provided
charge and mass resolution up to Be, allowing the final states of
interest to be unambiguously identified. The position and energy
resolution of the telescopes was $\sim$1 mm and 300 keV,
respectively. Calibration of the detectors was performed using
elastic scattering measurements of $^{9}$Be from $^{197}$Au and
$^{12}$C targets. The four telescopes were arranged in a
cross-like arrangement, separated by azimuthal angles of 90$^{\circ}$.
Two opposing detectors were located with their centres at
17.3$^{\circ}$ and 17.8$^{\circ}$ (telescopes 1 and 2) from the
beam axis and with the strip detector 130 mm from the target,
covering angular range from $\sim$7$^{\circ}$ to $\sim$28$^{\circ}$.
The remaining pair were at the slightly larger angles of 28.6$^{\circ}$
and 29.7$^{\circ}$ (telescopes 3 and 4), 136 mm from the target and
these telescopes covered angles from  $\sim$20$^{\circ}$ to
 $\sim$38$^{\circ}$. In the data acquisition system singles events were 
suppressed by factor 1000 and coincident events between any pair of 
telescopes were recorded.

\section{Results}

\subsection{$^{11}$C}

\begin{figure}
\includegraphics[width=1.0\textwidth]{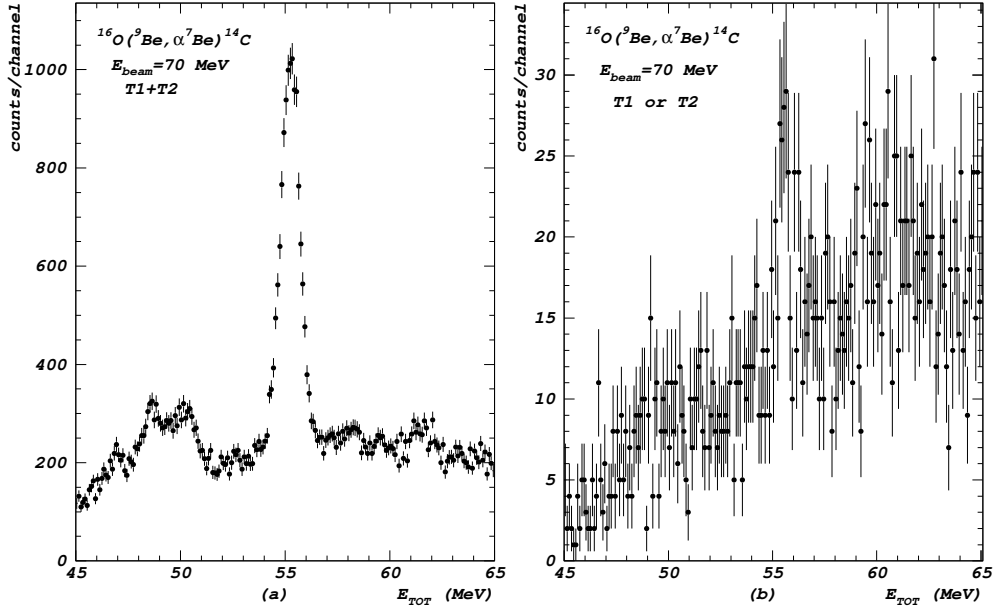}
\caption{\label{etot11c}
Total energy spectra for the $^{4}$He+$^{7}$Be (a) coincidences detected
in telescopes T1 and T2, (b) events when both particles are detected in
the same telescope T1 or T2. Statistical uncertainties are presented only.}
\end{figure}

The $\alpha$-decay of excited states in $^{11}$C has been studied using the
$^{16}$O($^{9}$Be, $^{11}$C* $\rightarrow$ $\alpha$+$^{7}$Be)$^{14}$C reaction
(Q = -14.602 MeV). The $^4$He and $^{7}$Be loci are well resolved in the particle 
identification spectra and the measurement of the energies and angles of these
detected particles permitted the kinematics of the reaction to be fully
reconstructed. Figure \ref{etot11c} shows the spectra for the total energy in
the reaction (a) for events when decay products were coincidently detected in 
telescopes T1 and T2 and (b) in the case when both $\alpha$ and $^{7}$Be were 
detected in the same telescope T1 or T2, assuming a $^{14}$C recoil. The
strongest peak in the spectrum shown in Fig. \ref{etot11c}(a) at E$_{tot}$=55.4 MeV 
corresponds to the $^{16}$O($^{9}$Be,$\alpha$$^{7}$Be) reaction. The energy 
resolution in this spectrum is 1.3 MeV, hence the contributions from the 
$^{7}$Be ground state and the first excited state, separated 429 keV, are 
unresolved. The three less intense peaks at lower total energy correspond to 
the $^{12}$C($^{9}$Be,$\alpha$$^{7}$Be)$^{10}$Be reaction (E$_{tot}$=50.5 MeV), 
the reaction on $^{16}$O when the recoiled $^{14}$C nucleus is excited to the second
excited J$^{\pi}$=0$^{+}$ state at 6.5894 MeV (E$_{tot}$=48.8 MeV) and the reaction 
on $^{12}$C target when the undetected $^{10}$Be is excited to the first excited 
state (E$_{tot}$=47.1 MeV).
There is small peak at the proper total energy (E$_{tot}$=55.4 MeV) also in the 
spectrum shown in Fig. \ref{etot11c}(b), but the background contribution
in this case is comparable to the reaction contribution. 
These events, double hits in T1 or T2, predominantly correspond to the small 
relative energy between the fragments and therefore to the excitations in 
$^{11}$C* close to the decay threshold.

\begin{figure}
\includegraphics[width=1.0\textwidth]{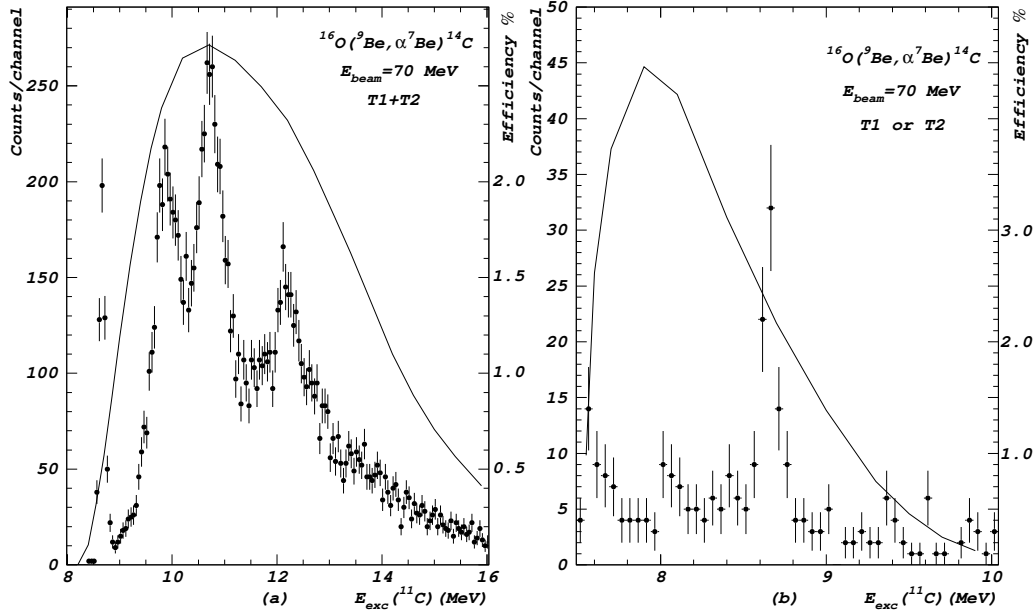}
\caption{\label{exc11c} $^{11}$C excitation energy spectra (a) for decays
detected in telescopes T1 and T2 in coincidence and (b) for the events
when both particles were detected in the same telescope T1 or T2. Error
bars represent statistical errors. The curves represent $\alpha$+$^{7}$Be 
detection efficiency calculated for the particular setup of detectors 
(right scale).}
\end{figure}

By selecting only the events associated with the peaks in the total energy 
spectra corresponding to the $^{16}$O($^{9}$Be,$\alpha$$^{7}$Be)$^{14}$C 
reaction it is possible to reconstruct the $^{11}$C excitation energy from
the relative velocity of the two decay fragments. Figure \ref{exc11c}
shows the $^{11}$C excitation energy spectra for (a) the events detected in 
the two opposing telescopes (T1+T2) and for (b) double hits in the same
telescope (T1 or T2). However, given that there are three particles in the 
final state, it is possible that they arise from decays of either $^{18}$O
into $\alpha$+$^{14}$C or $^{21}$Ne into $^{7}$Be+$^{14}$C. Both of
these possibilities were reconstructed and it is clear that there are no
contributions from either of these decay processes.
Figure \ref{exc11c} also shows results of the detection 
efficiency calculations for the specific detection geometry performed 
using Monte Carlo simulations in which isotropic $^{11}$C production and 
decay was assumed. The spectrum in Fig.\ref{exc11c}(a) is shown 
normalised for the detection efficiency in Fig. \ref{excs}(b). 
The excitation energy range covered for the
coincidence events in T1+T2 is from 8.4 to 17.5 MeV and for the double
hit events is from the threshold energy for the $\alpha$+$^{7}$Be decay
(E$_{thr}$=7.543 MeV) up to 10.5 MeV. The experimental excitation energy 
resolution is calculated to be 200-300 keV, close to the values found 
for the $\alpha$+$^{9}$Be \cite{so13c} and $\alpha$+$^{10}$Be 
\cite{so14c} events in the same data. The uncertainty in the excitation 
energy is 100 keV. The excitation energy spectrum in
Fig. \ref{exc11c}(a) shows strong peaks at 8.65, 9.85, 10.7 and 12.1
MeV and there is some evidence for additional peaks at 12.6 and
13.4 MeV. We note that no further peaks could be found in other telescope 
combinations. The spectrum in Fig. \ref{exc11c}(b) provides evidence for a
peak at 8.65 MeV. The observed states are presented in Table \ref{tab11cexc}. 
Given that the decays to the $^{7}$Be ground and the first 
excited state are unresolved in the total energy spectra, both processes 
may contribute to the $^{11}$C excitation energy spectra. These 
may be resolved in a two dimensional total energy versus $^{11}$C excitation 
energy spectrum. The peaks in the excitation spectrum corresponding to 
decays to the first excited state in $^{7}$Be should appear at the lower 
energy side of the peak in total energy spectrum and would lie 
429 keV below the peaks from the decays to the ground state.
Such satellite peaks are not observed and main peaks observed in the
$^{11}$C excitation energy spectra spread over the entire range of the reaction 
peak in the total energy spectrum. In other words, identical excitation 
energy spectra were obtained gating on the lower and higher energy side
of the reaction peak in the total energy spectrum. Consequently, this 
analysis provides evidence for only $\alpha$+$^{7}$Be(gs) decay of 
$^{11}$C states. However, a weak contribution of the $\alpha$+$^{7}$Be*(1/2$^{-}$)
decay cannot be excluded and is probably obscured in our data by the
more intense decay to the $^{7}$Be ground state. Because of the
considerable background for events of double hits in the same telescope, 
the excitation energy spectra were reconstructed for events
below and above the reaction peak in the total energy spectrum.
These spectra corresponding to the background events showed no evidence 
for the 8.65 MeV peak.

 The coincident detection of decay fragments in T1+T2 corresponds mainly to 
$^{11}$C* emission at forward angles up to 20$^{\circ}$ in the 
centre-of-mass system, there are almost no events for centre-of-mass emission 
angles greater than 30$^{\circ}$. In the case of double hits in the
same telescope (T1 or T2) the $^{11}$C* emission angle is larger than 
in the former case, the largest part of the events corresponds to the 
centre-of-mass angles between 20$^{\circ}$ and 35$^{\circ}$. The 
number of events in this latter case is much smaller, by a factor of 40 
(see Fig. \ref{etot11c}), which cannot 
be explained by changes in the detection efficiency (see Fig. \ref{exc11c}) or
in the reaction kinematics. Comparing the two spectra in Fig. \ref{etot11c} 
and considering the fact that the number of T3+T4 coincidences is much smaller 
than the number of T1+T2 events and that there are no double hits in T3 
and T4 telescopes, as well as the decrease of the T1+T2 yield with 
$^{11}$C* emission angle, it is evident that the reaction cross section 
decreases rapidly with increasing $^{11}$C* emission angle. This suggests 
that the main reaction mechanism for the population of $^{11}$C states was 
two-proton pickup to the cluster nucleus $^{9}$Be. The other possible direct
process, $^{5}$He knockout from $^{16}$O (or two step $\alpha$+n transfer) 
is very unlikely because the $^{5}$He spectroscopic factor in $^{16}$O 
is very small and such a process would enhance reaction cross section 
at larger $^{11}$C* emission angles. 

 An analysis of the angular distributions and
angular correlations for the states observed in Fig. \ref{exc11c}
using the techniques given in Ref. \cite{freac} was performed, but
these were found to be featureless. This is a consequence of the number
of reaction amplitudes contributing to the reaction process due to the
presence of nonzero spin nuclei in both the entrance and exit channels.

\subsection{$^{11}$B}

\begin{figure}
\includegraphics[width=1.0\textwidth]{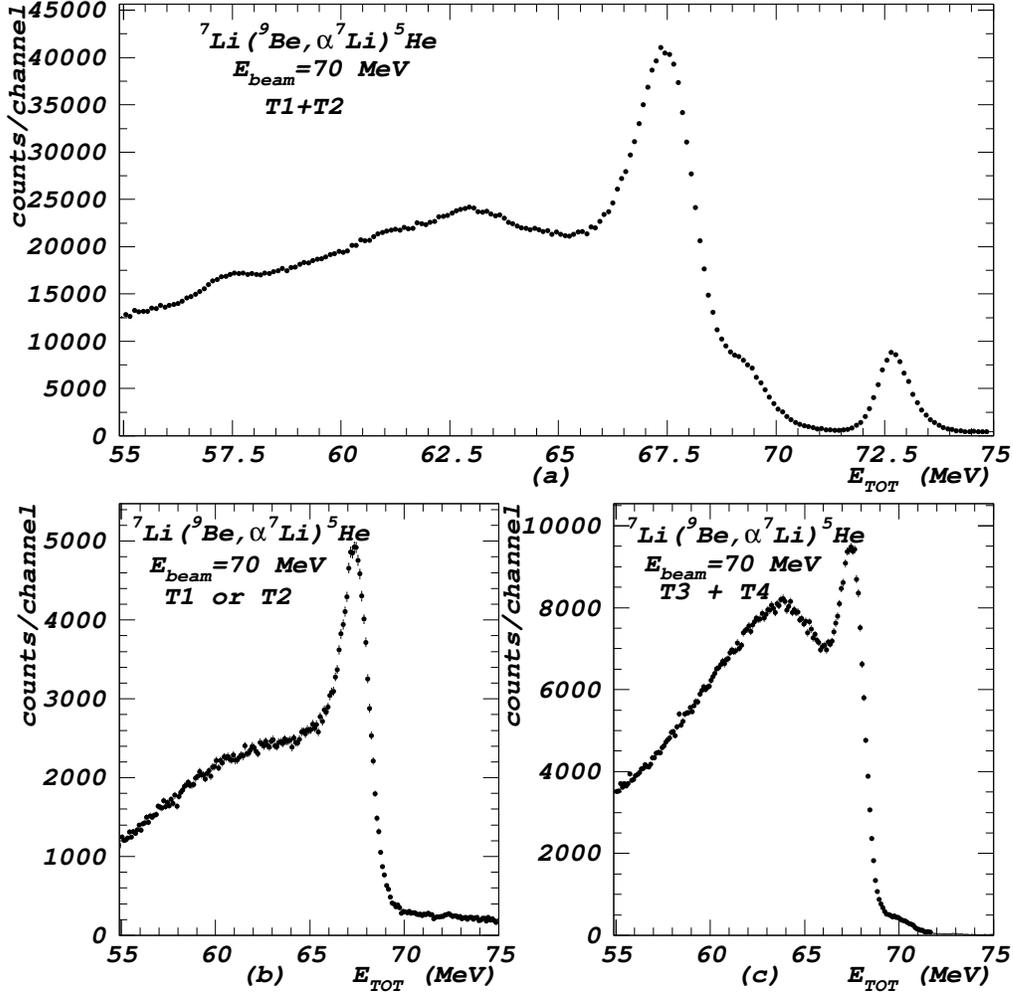}
\caption{\label{etot11b}
Total energy spectra for the $^{4}$He+$^{7}$Li (a) events coincidently 
detected in telescopes T1 and T2, (b) events when both particles are 
detected in the same telescope T1 or T2 and (c) coincident events in
telescopes T3 and T4. }
\end{figure}

\begin{figure}
\includegraphics[width=1.0\textwidth]{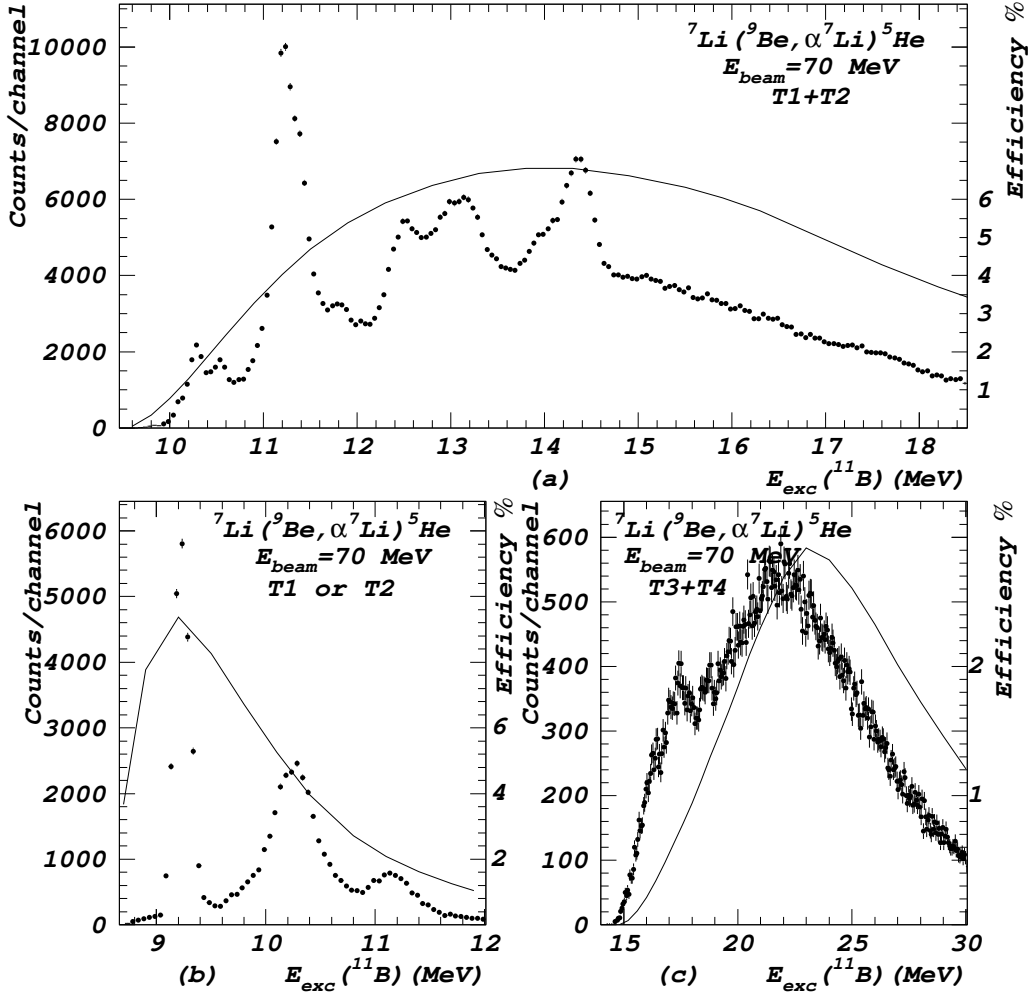}
\caption{\label{exc11b} $^{11}$B excitation energy spectra (a) for events
detected in telescopes T1 and T2 in coincidence, (b) for the events
when both particles were detected in the same telescope T1 or T2 and
(c) for the decays detected in T3 and T4. Error bars represent statistical 
errors. The curves represent $\alpha$+$^{7}$Li detection efficiency 
calculated for the particular setup of detectors (right scale).}
\end{figure}

A study of the $\alpha$-decay of $^{11}$B excited states has been performed 
using the $^{7}$Li($^{9}$Be,$^{11}$B* $\rightarrow$ $\alpha$+$^{7}$Li)$^{5}$He
reaction (Q = -2.4 MeV). The $^{7}$Li locus is resolved in the particle
identification spectra and only partially overlaps with the intense $^{6}$Li 
locus. The total energy spectra for this reaction are shown in Fig. 
\ref{etot11b}(a) for coincident events where decay products were detected 
in telescopes T1 and T2, (b) in the case when both $\alpha$ and $^{7}$Li were
detected in the same telescope T1 or T2 and (c) for coincident detection in
telescopes T3 and T4. The most intense peak in
these spectra, at E$_{tot}$=67.5 MeV, corresponds to the above reaction. Due to 
the experimental resolution and width of the $^{5}$He ground state (600 keV),
the contributions from the $^{7}$Li ground and the first excited state (separated 
478 keV) are unresolved and the widths of the peaks in the total energy 
spectra are $\sim$2 MeV. The highest energy peak at 72.6 MeV seen in Fig. 
\ref{etot11b}(a) arises from $\alpha$+$^{6}$Li coincidences from the very strong
$^{1}$H+$^{9}$Be $\longrightarrow$ $\alpha$+$^{6}$Li reaction leaking through the 
$^{7}$Li selection windows. The $\alpha$+$^{6}$Li*(3.56 MeV, T=1) events from the
same reaction produce a shoulder on the reaction peak around E$_{tot}$=69 MeV.
The threshold for the $^{7}$Li+$^{9}$Be $\longrightarrow$ 2$\alpha$+$^{7}$Li+n
channel is at 68.4 MeV and both the $^{5}$He* $\rightarrow$ $\alpha$+n and
$^{8}$Li* $\rightarrow$ $^{7}$Li+n processes may contribute at lower total
energy. A broad bump with a maximum around 63 MeV results from the contributions
of very broad $^{5}$He*(1/2$^{-}$) state and broad $^{8}$Li states between
5 and 10 MeV and also from the $^{7}$Li($^{9}$Be,$\alpha$$^{6}$Li) events.
This is also seen in Fig. \ref{etot11b}(c). There is also another bump in spectrum
(a) around 57.5 MeV arising from the ($^{9}$Be,$\alpha$$^{7}$Li) reaction on 
the oxygen component in the target.

By gating on the reaction peak in the total energy spectra, and selecting only 
the events associated with the $\alpha$+$^{7}$Li+$^{5}$He exit channel, $^{11}$B
excitation energy spectra can be reconstructed. Figure \ref{exc11b} shows 
such spectra for (a) the small angle detector pair events (T1+T2), (b) double
hit events in T1 or T2, and (c) for the larger angle pair events (T3+T4).
Again, no additional information on the $^{11}$B excitation energy spectrum was 
obtained from other telescope combinations. 
The three-body final state can be also produced via the decay of either $^{9}$Be
into $\alpha$+$^{5}$He or $^{12}$B into $^{7}$Li+$^{5}$He. The reconstructed
$^{12}$B excitation energy spectra show no evidence for the $^{7}$Li+$^{5}$He
decay. In addition, the $^{9}$Be excitation energy spectra for the T1+T2 events and
double hit events in T1 or T2 show that there is no contribution from the 
$\alpha$+$^{5}$He decay for these detection geometries. The $^{9}$Be excitation 
spectra for T3+T4 events show evidence for the known $\alpha$+$^{5}$He decay of 
the excited states at 2.4, $\sim$6.5 and $\sim$11.5 MeV which was also observed in 
\cite{so9be}. The contributions from the $^{9}$Be states below 8.6 MeV in excitation 
were removed from the spectrum in Fig. \ref{exc11b}(c) but a weak contribution from 
the 11.5 MeV state, and possible higher states, remains. 
The results of the detection efficiency calculations for the $\alpha$+$^{7}$Li decay 
of $^{11}$B are also presented in Fig. \ref{exc11b}. Again, the spectrum in Fig.
\ref{exc11b}(a) is shown in Fig. \ref{excs}(a) normalised for the variation in 
detection efficiency. The $^{11}$B excitation 
energy spectra extend from the threshold at 8.664 MeV (Fig. \ref{exc11b}(b)) to
$\sim$30 MeV (Fig. \ref{exc11b}(c)). The uncertainty in the excitation energy is 
100 keV, and the experimental excitation energy
resolution for the low excitations is again calculated to be 200-300 keV. The
spectrum in Fig. \ref{exc11b}(a) shows clear resonances at 10.3, 10.55, 11.2,
11.8, 12.5, 13.1 and 14.35 MeV and there are indications for additional peaks at
11.4, 13.0, 14.0 and $\sim$17.4 MeV. In the spectrum for the events in the 
same telescope, Fig. \ref{exc11b}(b),  peaks are observed at 9.2, 10.25
and 11.15 MeV. The 9.2 MeV state is also observed in the T3+T3 and T4+T4 data. 
The spectrum for the T3+T4 events, Fig. \ref{exc11b}(c), indicates a
state at 17.4 MeV and a weak state at 18.6 MeV.
The observed states are presented in Table \ref{tab11bexc}.
These spectra contain possible contributions from both the decays to the ground 
and the first excited state of $^{7}$Li. An analysis identical to the one 
performed for the $^{11}$C data, gives evidence for only $\alpha$+$^{7}$Li(gs)
decay of the $^{11}$B excited states. Thus, the much weaker 
$\alpha$+$^{7}$Li*(1/2$^{-}$)
decay is again hidden in our data by the dominant decay into $^{7}$Li(gs).

The main reaction mechanism for the population of $^{11}$B excited states when 
decay fragments were detected at forward angles (T1+T2 coincidences for which the 
$^{11}$B* centre-of-mass emission angle was mainly less than 20$^{\circ}$) is 
expected to be deutron (or n+p) pickup from $^{7}$Li to $^{9}$Be. At larger $^{11}$B*
emission angles (for the largest part of T3+T4 coincidences centre-of-mass emission 
angle was between 10$^{\circ}$ and 30$^{\circ}$ and for double hits in T1 or T2 it 
is mainly between 30$^{\circ}$ and 60$^{\circ}$) contributions from $\alpha$-transfer 
from $^{9}$Be to $^{7}$Li are also possible.

The analysis of the angular distributions and angular correlations were performed 
for the main states observed in Fig. \ref{exc11b} but these were found to be 
featureless and did not give any information about the spin of the states.
The angular distributions for the detection of $^{11}$B* in T1+T2 showed a 
strong peak at very forward angles and then a decrease in yield with increasing
$^{11}$B* angle. These characteristics are consistent with the direct reaction 
mechanism.

\section{Discussion}

\subsection{$^{11}$C}

\begin{table}
\caption{\label{tab11cexc} $^{11}$C excited states decaying into $\alpha$+$^{7}$Be(gs) 
from the present measurement, the previous measurements of the 
$^{6}$Li($^{10}$B,$\alpha$$^{7}$Be) \cite{lee} and $^{7}$Be($\alpha$,$\gamma$) 
\cite{har84} reactions and known levels at these excitations from 
the tabulations of Ref. \cite{ajz90}. 
The uncertainty in the excitation energy of the present measurement is 100 keV.   }

\begin{tabular}{|c|c|c|c|c|c|c|}\hline
\multicolumn{1}{|c|}{Present} & \multicolumn{1}{|c|}{Ref. \cite{lee} } & 
\multicolumn{1}{|c|}{Ref. \cite{har84} } & \multicolumn{4}{|c|}{Tabulations \cite{ajz90} }\\
\hline
$E_x$ (MeV) & $E_x$ (MeV) & $E_x$ (MeV) & $E_x$ (MeV)  &  Width (keV) & $J$ ; T & Reference \\
\hline

       & 8.10  & 8.105 & 8.1045 & 11 eV    & 3/2$^-$   & \cite{fuchs,smith}\\
       & 8.42  & 8.421 & 8.420  & 15 eV    & 5/2$^-$   & \cite{fort73,fuchs,wie83,tonch,cosp,amann,smith}\\
8.65   & 8.655 &       & 8.655  & $\leq$ 5 & 7/2$^+$   & \cite{fort73,fuchs,wie83,amann,smith} \\
       &       &       & 8.699  & 15       & 5/2$^+$   & \cite{fort73,ang,fuchs,wie83,tonch,amann,smith} \\
       &       &       & 9.20   & 500      & 5/2$^+$   & \cite{wie83}\\
       &       &       & 9.65   & 210      & (3/2$^-$) & \cite{wie83,tonch}\\
9.85   &       &       & 9.78   & 240      & (5/2$^-$) & \cite{fuchs,wie83,tonch,ove,cro}\\ 
       &       &       & 9.97   & 120      & (7/2$^-$) & \cite{wie83,smith}\\ 
       &       &       & 10.083 & 230      & 7/2$^+$   & \cite{fuchs,wie83,ove,cro}\\
10.7   &       &       & 10.679 & 200      & 9/2$^+$   & \cite{fuchs,wie83,jen64,ove,smith}\\
       &       &       & 11.03  & 300      &           & \cite{fuchs,bene}\\
       &       &       & 11.44  & 360      &           & \cite{jen64,rihe}\\
12.1   &       &       & 12.16  & 270      & T=3/2     & \cite{wats}\\
       &       &       & 12.4   & 1-2 MeV  & $\pi$=-   & \\
       &       &       & 12.51  & 490      & 1/2$^-$; 3/2 & \cite{wats,bril,cosp,bene,macdo,amann}\\ 
(12.6) &       &       & 12.65  & 360      & (7/2$^+$) &  \cite{jen64,bene}\\
       &       &       & (13.01)&          &           & \\
       &       &       & 13.33  & 270      &           & \cite{jen64,bene}\\
(13.4) &       &       & 13.4   & 1100     &           & \cite{smith}\\

\hline
\end{tabular}
\end{table}

The $^{11}$C excited states observed in the present measurement as well as
the information about $\alpha$-decaying states from previous measurements
\cite{lee,har84} and known levels from the tabulation 
of the $^{11}$C states \cite{ajz90} are presented in Table \ref{tab11cexc}.
The threshold energy for the $\alpha$+$^{7}$Be, p+$^{10}$B,
2$\alpha$+$^{3}$He, $^{8}$Be+$^{3}$He and n+$^{10}$C decays are at 7.543,
8.6896, 9.131, 9.223 and 13.120 MeV respectively. Our results show a number
of $\alpha$-decaying states above the proton threshold energy, which may be
an indication of their $\alpha$-cluster structure. 

The only published coincidence measurement of the $^{11}$C* $\rightarrow$ 
$\alpha$+$^{7}$Be decay performed using the 
$^{6}$Li($^{10}$B,$\alpha$$^{7}$Be)$^{5}$He reaction at E$_{beam}$=65 MeV
\cite{lee} reports states at 8.10, 8.42 and 8.655 MeV also decaying only 
to the $^{7}$Be ground state. In both this and the present measurement, 
the contributions to the peak at 8.65 MeV may come from the very narrow states 
at 8.655 (7/2$^+$) and 8.699 MeV (5/2$^+$) \cite{fort73}. It is worth 
mentioning that the 5/2$^+$ state, which is only 10 keV above the threshold
for proton decay, strongly enhances the cross section of the astrophysically 
important $^{10}$B(p,$\alpha$) reaction \cite{ang}.  
We do not observe the 8.10 (3/2$^-$) and 8.42 MeV (5/2$^-$) states in 
our spectra.  These two negative-parity states were observed
in the only published study of the $^{7}$Be($\alpha$,$\gamma$) reaction 
\cite{har84} which provided following results: $\Gamma_{\gamma}$=0.35 eV,
$\Gamma_{\alpha}$=4-18 eV (most probable value is 6 eV) for the 8.1045 MeV
state and $\Gamma_{\gamma}$=3 eV, $\Gamma_{\alpha}$=13 eV for the 8.42 MeV
state. Their very small widths for the $\alpha$-decay may be understood to be
a consequence of the Coulomb barrier and also the centrifugal barrier in the 
case of the 5/2$^-$ state for which decay proceeds with L=2. 
 
From Fig. \ref{exc11c}(b) it is evident that the detection efficiency 
for these lower excitations is higher than for the 8.65 MeV state and 
absence of these two states in the present
spectrum cannot be a consequence of the detection geometry. One possible 
explanation is that the different $^{11}$C* population mechanisms 
play a role, i.e. single proton transfer to $^{10}$B(3$^+$) in Ref. \cite{lee} 
and two proton transfer to $^{9}$Be(3/2$^-$) in the present case.
But, given that $^{10}$B ground state corresponds to p$_{3/2}$ proton coupled to 
the $^{9}$Be(gs) core, these two reactions are expected to be quite similar.

Simple semiclassical considerations, which assume that the transfer 
cross-section would be large when the velocities of the incident and 
final nuclei are the same, shows that the kinematics of the present 
measurement prefers population of $^{11}$C states with transferred value 
L=3-4, while the measurements in Ref. \cite{lee} prefer L=1-2 transfers. 
It should be mentioned that the results from the measurements of the 
$^{9}$Be($^{3}$He,n)$^{11}$C reaction \cite{fuchs} at E$_{beam}$=10.5 and 
13 MeV (this kinematics prefers L=2 transfer of two protons) provided also 
clear evidence for the 3/2$^-$, 5/2$^-$ states. 
From the angular distributions obtained in these measurements, it was
concluded that the 8.105 MeV state structure should be 
(p$_{3/2}^{5}$)$_{3/2}$p$_{1/2}^{2}$, the 8.42 MeV state has structure 
(p$_{3/2}^{6}$)$_{3}$p$_{1/2}$, the 5/2$^+$, 7/2$^+$ states
at 8.7 MeV show large components of s$_{1/2}$ and also d$_{5/2}$
particles added to (p$_{3/2}^{6}$)$_{3}$
and the most probable configuration of the 10.679 MeV state is
(p$_{3/2}^{6}$)$_{3}$d$_{5/2}$.

The next peak in the present spectra at 9.85 MeV corresponds to
at least two known $\alpha$-decaying states at 9.65 and 10.083 MeV, 
but contributions of the all four known states between 9.6 and 10.1 MeV 
are possible. The list of levels between 8 and 11 MeV accepted in the
tabulations of Ref. \cite{ajz90} was established in a detailed study
of the $^{10}$B(p,$\gamma$)$^{11}$C reaction \cite{wie83} which 
found three negative-parity states for excitations between 9.6 
and 10.0 MeV. A recent measurement of the astrophysically important 
$^{10}$B($\vec{p}$,$\gamma$)$^{11}$C reaction at low energy \cite{tonch} 
has shown that the two p-wave resonances at 9.65 and 9.78 MeV, as well 
as the 8.420 MeV J$^{\pi}$=5/2$^-$ subthreshold state, also contribute 
to the capture process.

The 10.7 MeV peak in the present spectrum corresponds 
to the known 9/2$^+$ $\alpha$-decaying state at 10.679 MeV.
The 10.67, 12.65 and 13.33 MeV states were observed in the measurement
of the $^{10}$B(p,$\alpha$)$^{7}$Be(gs) reaction \cite{jen64} which also
provided evidence for the 11.44 MeV state in $^{7}$Be*(1/2$^-$) channel
and for some states at higher excitations. The other measurements of the
same reaction report states at 10.09, 10.68 MeV \cite{ove} and 
9.76, 10.06 MeV \cite{cro}. Absence of the 11.03 MeV state, which
is populated in two proton transfer reaction \cite{fuchs}, in our
spectra provides evidence for its very small $\alpha$ width and 
preferential decay by proton emission. The $\alpha$+$^{7}$Be*(1/2$^-$)
decay of the 11.44 MeV state was also observed in the measurements of
the $^{10}$B(p,$\alpha$$\gamma$) reaction \cite{rihe}. Evidently, it
does not decay, or decays weakly, into $^{7}$Be(gs) channel which may 
imply that its spin is 1/2$^{-}$.

The unexpected result observed in the present data is the strong 
$\alpha$+$^{7}$Be(gs) decay of the 12.1 MeV state (which means T=1/2),
which is believed to be the isobaric analogue state of the T=3/2 
$^{11}$Be ground state. This value of isospin has been accepted in the 
tabulations of the $^{11}$C properties \cite{ajz90}, but comes from only 
one published experimental work \cite{wats}. These measurements of the
$^{11}$B($^{3}$He,t), $^{9}$Be($^{3}$He,n) and 
$^{10}$B(p,p')$^{10}$B*(1.74 MeV, T=1) reactions gave very tentative 
evidence for isobaric analogue states of the three lowest $^{11}$Be 
states at 12.17, 12.57 and 13.92 MeV. The observed weak peaks attributed 
to the 12.17 MeV state are questionable in the spectra of all these 
reactions, which populate states of both isospin values 1/2 and 3/2. 
In the other measurement of the $^{9}$Be($^{3}$He,n) reaction \cite{bril} 
levels observed at 12.5, 13.7 and 14.7 MeV were tentatively identified 
as the analogs of the three lowest excited states in $^{11}$Be. 
Measurements of the $^{13}$C(p,t) reaction \cite{cosp,bene,macdo} 
provided evidence for the T=3/2, J$^{\pi}$=1/2$^{-}$ state at 12.47 MeV 
with total width of 500 keV which is the analog of the first excited
state in $^{11}$Be. This state was also observed in the 
$^{12}$C($\pi^{+}$,p) but not in the $^{12}$C(p,d) reaction which
confirms its T=3/2 character \cite{amann}. 

It seems from these results that the 1/2$^{-}$ and even 5/2$^{+}$ state 
have been identified experimentally, but the status of the 1/2$^{+}$ state 
is unclear. Our result provides evidence for a previously unobserved T=1/2 
state at 12.1 MeV whose width is comparable to the width of the 10.679 MeV 
state, which is 200 keV (the peaks at 10.7 and 12.1 MeV in Fig. \ref{exc11c}(a) 
have the same experimental width of 400 keV). This state may be the same 
state observed in Ref. \cite{wats}. We should emphasize that, except in 
Ref. \cite{wats}, the present result is the only observation of the state in 
this excitation region (11.4-12.4 MeV). If this state is really the isobaric 
analogue state of the $^{11}$Be ground state, as was claimed in 
\cite{wats,ajz90}, it has a very strong and unexpected isospin mixing. A 
simpler explanation may be that the 12.1 MeV level is indeed T=1/2 and
possesses an alternative (rotational) structure, and the true
(1/2$^{+}$,3/2) state in $^{11}$C has not yet been identified
experimentally. We return to this point later (section 4.3).

\subsection{$^{11}$B}

\begin{table}
\caption{\label{tab11bexc} $^{11}$B excited states decaying into 
$\alpha$+$^{7}$Li(gs) from the present measurement and known unbound states
below 19 MeV from the tabulations of Ref. \cite{ajz90}.  The uncertainty in the 
excitation energy of the present measurement is 100 keV.   }

\begin{tabular}{|c|c|c|c|c|}\hline
\multicolumn{1}{|c|}{Present} &  \multicolumn{4}{|c|}{Tabulations \cite{ajz90} }\\
\hline
 $E_x$ (MeV) & $E_x$ (MeV)  &  Width (keV) & $J$; T & Reference \\
\hline
        
        &    8.9202    & 4.37 eV &  5/2$^-$   & \cite{har84,cosp,macdo,zwieg,arya}\\ 
 9.2    &    9.1850    &   2 eV  &  7/2$^+$   & \cite{har84,schmidt,zwieg}\\
        &    9.2744    &   4     &  5/2$^+$   & \cite{har84,paul,schmidt,zwieg}\\
        &    9.82      &         & (1/2$^+$)  & \\
        &    9.876     & 110     &  3/2$^+$   & \cite{cuss,paul,schmidt,zwieg}\\
 10.3   &   10.26      & 150     &  3/2$^-$   & \cite{cuss,paul,schmidt,jambu,zwieg,arya}\\
        &   10.33      & 110     &  5/2$^-$   & \cite{cuss,paul,schmidt,jambu,zwieg,arya}\\
 10.55  &   10.597     & 100     &  7/2$^+$   & \cite{cuss,paul,schmidt,haus,zwieg,arya}\\
        &   10.96      & 4500    &  5/2$^-$   & \cite{cuss,schmidt}\\ 
 11.2   &   11.265     & 110     &  9/2$^+$   & \cite{cuss,schmidt,fletcher,zwieg}\\
(11.4)  &   11.444     & 103     &            & \cite{cuss,paul,schmidt,jambu,zwieg}\\
        &   11.600     & 170     &  5/2$^+$   & \cite{macdo,cuss,schmidt,haus,zwieg}\\
 11.8   &   11.886     & 200     &  5/2$^-$   & \cite{cuss,haus,zwieg}\\
        &   12.0       & $\sim$1000 &  7/2$^+$ & \cite{cuss}\\
 12.5   &   12.557     & 210     &  1/2$^+$ (3/2$^+$); 3/2 & \cite{wats,cuss,jambu,fletcher,goos,zwieg}\\
(13.0)  &   12.916     & 200     &  1/2$^-$; 3/2 & \cite{wats,cosp,bene,macdo,goos,zwieg,arya}\\
 13.1   &   13.137     & 426     &  9/2$^-$   & \cite{cuss,haus,zwieg}\\
        &   13.16      & 430     &  5/2$^+$,7/2$^+$ & \cite{jambu,haus}\\
(14.0)  &   14.04      & 500     &  11/2$^+$  & \cite{cuss,jambu,haus,goos}\\
14.35   &   14.34      & 254     &  5/2$^+$; 3/2 & \cite{wats,jambu,fletcher,goos,zwieg}\\ 
        &   14.565     & $\leq$30 &           & \cite{cuss,zwieg,arya}\\
        &   15.29      & 250     &  (3/2,5/2,7/2)$^+$;(3/2) & \cite{haus,goos,arya}\\
        &   16.437     &  $\leq$30 & T=3/2    & \cite{zwieg,zwieg75,arya}\\ 
        &   17.33      &  $\sim$1000 &        & \\
(17.4)  &   17.43      & 100     &  T=3/2     & \cite{zwieg,zwieg75}\\ 
        &   18.0       & 870     &  T=3/2     & \cite{zwieg}\\ 
(18.6)  &   18.37      & 260     &  (1/2,3/2,5/2)$^+$ & \cite{zwieg75}\\ 

\hline
\end{tabular}
\end{table}

Table \ref{tab11bexc} presents the observed $\alpha$-decaying $^{11}$B 
excited states in the present measurement and known states from the tabulations 
of Ref. \cite{ajz90}. The threshold energies for the $\alpha$+$^{7}$Li,
2$\alpha$+t, $^{8}$Be+t, p+$^{10}$Be, n+$^{10}$B and d+$^{9}$Be decays 
are at 8.664, 11.131, 11.223, 11.228, 11.454 and 15.815 MeV, respectively.
Once again, the observed $\alpha$-decay of states above the thresholds for other
decay channels may indicate the $\alpha$-cluster structure of these states.

The lowest energy peak in our $^{11}$B excitation spectra (Fig. \ref{exc11b}) 
appears at 9.2 MeV and corresponds to the 7/2$^{+}$, 5/2$^{+}$ doublet of very 
narrow states at this excitation. Absence of the 8.9202 MeV 5/2$^{-}$ state,
which is only 256 keV above the threshold, can be explained
in terms of the Coulomb and centrifugal barrier (it decays with L=2).
These three states have been observed in the measurement of the 
$^{7}$Li($\alpha$,$\gamma$) reaction \cite{har84} in which  
radiative and $\alpha$-partial widths for these states were extracted. 
A very small $\alpha$ width for the 8.9202 MeV state was found 
($\Gamma_{\gamma} / \Gamma\approx 1$) and for the 9.185 MeV state it was found 
that $\gamma$ width is about 10$\%$ of the total width. 

The next peaks in the present spectra are at 10.3 and 10.55 MeV which correspond
to the 3/2$^{-}$ state at 10.26 MeV and 5/2$^{-}$ at 10.33 MeV, and the 10.597 MeV 
7/2$^{+}$ state. The strongest observed peak corresponds to the 11.265 MeV 9/2$^{+}$
state. Slight asymmetry in its shape at higher energy may be due to the 11.444
MeV state. The next weak peak is the 11.886 MeV 5/2$^{-}$ state. These states 
are all known as $\alpha$-decaying states from many studies.
  
The measurements of the $^{7}$Li+$\alpha$ elastic and inelastic 
scattering to the first excited state of $^{7}$Li \cite{cuss} found 
evidence for the states at 10.34, 10.60, 11.29, 11.49 and 12.55 MeV in the 
elastic channel and 9.88, 10.25, 10.60, 10.96, 11.29, 11.49, 11.60, 11.88 
and 12.55 MeV in the inelastic channel, and also indications of levels at
12.04, 13.03, 14.05, 14.69 and 15.79 MeV. A study of the 
$^{7}$Li($\alpha$,$\gamma$)$^{11}$B and 
$^{7}$Li($\alpha$,$\gamma$$\alpha$)$^{7}$Li*(0.478 MeV) reactions \cite{paul}
reported states at 9.28, 9.88, 10.26, 10.32 and 10.62 MeV and also an
indication of a level at 10.45 MeV. 
In a kinematically complete measurement of the $^{14}$N(n,2$\alpha$)$^{7}$Li
reaction \cite{schmidt} the $\alpha$+$^{7}$Li(gs) decay of the 
states at 9.19, 9.277, 10.25 and 10.60 MeV was observed.
Coincidence measurements of the $^{9}$Be+$^{6}$Li reaction \cite{jambu} 
provided evidence for the $\alpha$-decaying levels in $^{11}$B at 10.3, 11.4, 
12.6, 13.16, 13.5, 14.0 and 14.4 MeV. 

Our results show a peak at 13.1 MeV, which corresponds to both the 
13.137 MeV 9/2$^{-}$ and 13.16 5/2$^{+}$ (or 7/2$^{+}$) state \cite{haus}
and also indicate a weak broad state at 14.0 MeV which corresponds to
the 11/2$^{+}$ state at 14.04 MeV and there may be some indication of the 
state at 18.6 MeV in Fig. \ref{exc11b}(c).

A curious feature of the present results is the observation of the 
$\alpha$+$^{7}$Li(gs) decay, which means isospin T=1/2, of the excited states 
at 12.557, 12.916, 14.34 and 17.43 MeV proposed to be the isobaric analogue 
states of the $^{11}$Be states which have T=3/2. The 12.557 MeV state, 
which should be the analogue of the ground state, is strong in our spectra 
as well as 14.34 MeV 5/2$^{+}$ state. The widths of these states estimated 
from Fig. \ref{exc11b}(a) are in agreement with their accepted values 200-250 
keV \cite{ajz90}. The 12.916 1/2$^{-}$ state is close to the peak at 
13.1 MeV, but there is good evidence for an additional peak around 13.0 MeV in 
our spectra. The peak observed in Fig. \ref{exc11b}(c) at 17.4
MeV may correspond to the previously observed state at 17.43 MeV claimed to
be T=3/2 state. The 12.55 MeV state was also observed in the
$^{7}$Li+$\alpha$ scattering \cite{cuss} and the $^{9}$Be+$^{6}$Li reaction 
\cite{jambu} where the 14.4 MeV state was also observed. The resonances at
12.5 and 14.3 MeV (and also 11.3 MeV) have also been observed in recent 
measurement of the $^{7}$Li($^{7}$Li,$^{7}$Li$\alpha$) reaction 
\cite{fletcher}. 

Information about T=3/2 states in $^{11}$B has been obtained 
from a number of measurements of different reactions. A measurement of the 
$^{10}$Be(p,$\gamma$) reaction provided evidence for the states at
12.55, 12.91, 14.33 and 15.3 MeV which were identified as the analogues
of the lowest four states in $^{11}$Be \cite{goos}. 
A study of the $^{9}$Be($^{3}$He,p) and $^{9}$Be($\alpha$,d) reactions
\cite{zwieg} found evidences for the T=3/2 states at 12.56, 12.92, 14.47, 16.44,
17.69, 18.0, 19.15 and 21.27 MeV. It was concluded that the 16.44, 
17.69, 18.06 and 19.15 MeV 
states had a rather pure isospin 3/2, whereas the first two may have small
admixtures of T=1/2 since they were seen in the isospin-forbidden 
$^{9}$Be+d reaction \cite{zwieg75}. The 14.47 MeV state was suggested to 
correspond to the 14.33 MeV excitation and thus to
have a strongly mixed isospin because it appeared in the spectra from 
both reactions (detailed discussion of its properties was presented in 
Ref. \cite{zwieg}). We note that these observations are in good agreement 
with the present results. The 17.69 MeV state observed in the above work 
may be the 17.43 MeV state observed in the present measurements.

A measurement of the $^{14}$C(p,$\alpha$) reaction \cite{arya} showed the
population of broad resonances at 12.92, 15.29, 16.50 and 
19.07 MeV which were proposed to be T=3/2 negative parity states.
The 12.92 MeV state was also observed in the $^{13}$C(p,$^{3}$He) reaction 
\cite{cosp,bene,macdo}. 
From all the available data on T=3/2 states in $^{11}$B it seems that, at
least, the lowest states have been identified experimentally. 
However, the earlier measurements together with the present analysis points 
to a significant T=1/2 contribution.

We should note that as in the case of $^{11}$C there is an
alternative explanation of the some of the states observed in the
present measurement in terms of rotational bands.

\subsection{Common features of the $^{11}$B and $^{11}$C excited states}

As just indicated, there are two possible interpretations of the
present data. The first is that several of the observed states 
coincide with known T=3/2 states, in which case isospin mixing is 
signalled. Alternatively, the peaks may have a genuine T=1/2 
character and may be linked to rotational bands. We deal with
each of these possibilities in turn.

First the T=3/2 states in both nuclei will be examined. If the 
presented states are indeed those identified with T=3/2 character,
then our results show that the lowest three T=3/2 levels in $^{11}$B 
and the first T=3/2 level in $^{11}$C probably have large isospin mixing. 
In the $^{11}$B case this is confirmed by other published results of 
observations of the T=1/2 resonances at excitations proposed for the 
T=3/2 levels.

The results of several calculations \cite{fort95,sherr01} suggest that 
the 1/2$^{+}$, T=3/2 levels in $^{11}$B and $^{11}$C have been misidentified. 
But the latest published results \cite{barker} of potential-model 
calculations using more appropriate R-matrix definitions for the energy 
and width of an unbound level, and many-channel R-matrix theory, have found 
reasonable agreement with the experimental excitations but possible 
disagreement in the widths. It has been proposed that isospin mixing can 
resolve confusion with T=3/2 states in $^{11}$B and $^{11}$C. Shell-model 
calculations \cite{barker,teet,jager} suggest a 1/2$^{+}$, T=1/2 partner 
state near the 1/2$^{+}$, T=3/2 state. Also, both the shell-model 
\cite{cohen,cohen70,wolt} and three-cluster model 
calculations \cite{des} of $^{11}$B predict 1/2$^{-}$, T=1/2 state 
above the $\alpha$+$^{7}$Li decay threshold which may mix with the 
T=3/2 partner state, but this state has not been observed.
Two parentages appear in these states, wave functions of $^{11}$B T=3/2
states contain one third of $^{10}$Be$\otimes$p and two thirds of
$^{10}$B*(T=1)$\otimes$n while $^{11}$C states are two thirds of
$^{10}$B*(T=1)$\otimes$p and one third of $^{10}$C$\otimes$n. 
The weak coupling model calculations in a complete 1$\hbar\omega$ basis 
\cite{teet} showed that the T=3/2 $^{11}$B positive parity states wave
functions are quite simple, these states consist mainly of the proton 
in the sd-shell coupled to the $^{10}$Be ground state and components of 
the sd-shell proton coupled to the $^{10}$Be first excited state. There 
is the possibility for T=1/2 positive parity states at these excitations 
based upon the configurations with the ground state of $^{8}$Be as an 
inert core and three particles in (2s,1d) shell \cite{true} and also for 
negative parity states with two particles in the sd-shell, which
have large overlap with the $\alpha$+$^{7}$Li structure. These states
may then mix with the observed T=3/2 states of the same spin and parity.

\begin{figure}
\includegraphics[width=0.75\textwidth]{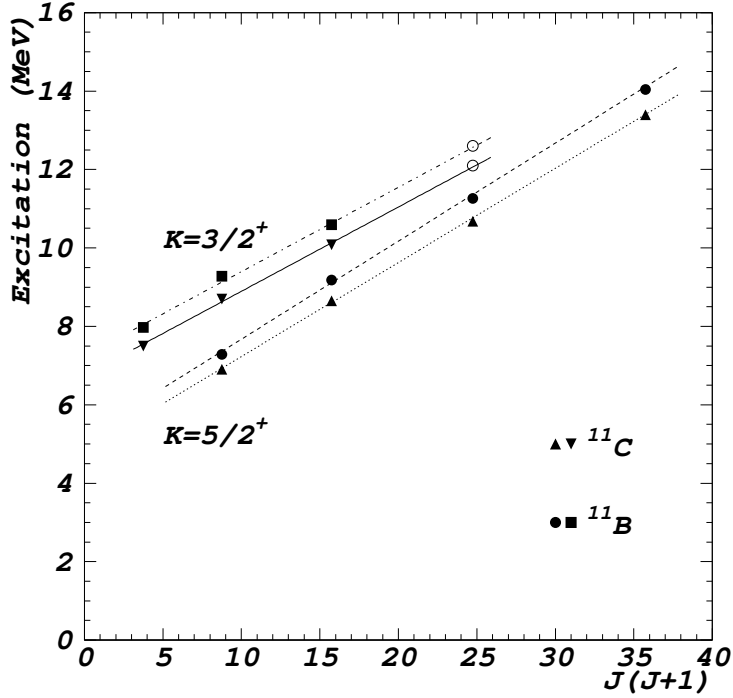}
\caption{\label{ejj}
The members of proposed positive-parity rotational bands in $^{11}$B and $^{11}$C.
Excitation energies of states are shown as function of J(J+1). The positions of
possible 9/2$^{+}$ states in K=3/2$^{+}$ bands are marked with open circles. }
\end{figure}

We now examine the possible rotational behaviour of the mirror nuclei.
An interesting result was obtained in the Nilsson-Strutinsky cranking model 
calculations for the positive-parity yrast states of $^{11}$C and $^{11}$B
\cite{ragn}. Based on available experimental data, rotational bands 
with K=5/2$^+$ were proposed beginning at 7.286 and 6.905 MeV in $^{11}$B and 
$^{11}$C respectively, with rotational members at 9.185 and 8.655 (7/2$^+$), 
11.265 and 10.679 (9/2$^+$) and 14.04 and 13.33 MeV (11/2$^+$) (see Fig. 
\ref{ejj}). The moment of inertia I of these bands is very large, with a 
rotational parameter $\hbar^{2}$/2I of 0.25 MeV for $^{11}$B and 0.24 MeV 
for $^{11}$C (for comparison rotational parameter for the $^{8}$Be ground state 
band is 0.5 MeV) which would correspond to an extremely deformed structure. Two 
alternative explanations were offered for these bands, the first was that there 
are three particles promoted to the sd-shell [220]1/2$^+$, but that was in 
conflict with the generally accepted signature selection rule. The second 
explanation, which seems more likely, was that the bands could be a 1p-1h 
excitation to the sd-shell, presumably the oblate coupled [202]5/2$^+$ orbit
leaving an unpaired neutron and proton in the p-shell.

What is interesting here, is that the present results show the population of 
the members of rotational bands in the reactions which involve two-nucleons
transferred to the cluster nucleus $^{9}$Be and their strong 
$\alpha$+$^{7}$Li($^{7}$Be) decay. This fact indicates that the configurations
of these states are rather complicated, because simple 1p-1h configuration
would strongly decay by single-nucleon emission. 

In $^{11}$B spectra all the 
members 7/2$^+$, 9/2$^+$ and 11/2$^+$ are present while in $^{11}$C spectra 
the 7/2$^+$ and 9/2$^+$ states are clearly seen and there is small bump in 
the $^{11}$C spectra at 13.4 MeV. The assigned 11/2$^+$ member in Ref. 
\cite{ragn}, the 13.33 MeV state with width of 270 keV observed in Ref.
\cite{jen64}, has no experimentally measured spin and parity. Comparing 
the widths of the known states of both nuclei in the compilations \cite{ajz90}, 
it seems that the broad state observed at 13.4 MeV is better candidate for 
that level. The excitation of this level is not unambiguously confirmed
and it may correspond to broad state observed in the $^{12}$C(p,d)
measurement \cite{smith} at 13.22 $\pm$ 0.25 MeV.

Interestingly, there is another possible positive-parity band in both
$^{11}$B and $^{11}$C with K=3/2$^+$, beginning at excitation energies
7.97784 and 7.4997 MeV, respectively. Rotational members at 9.274 and 8.699 
MeV (5/2$^+$) and 10.597 and 10.083 MeV (7/2$^+$), which are observed in the 
present spectra, form the linear J(J+1) energy plot presented in Fig. \ref{ejj}. 
The rotational parameter $\hbar^{2}$/2I of these two bands would be 0.215 MeV.

It should be noted that the 9/2$^+$ members of these bands (open circles in 
Fig. \ref{ejj}) would be at 12.6 MeV in $^{11}$B and 12.1 MeV in $^{11}$C, 
they are very close to the excitations of the proposed 1/2$^+$ T=3/2 states and 
exactly at excitations where are resonances in the present spectra. This offers 
an alternative explanation of these peaks in our spectra: they may be the 9/2$^+$ 
T=1/2 members of the very deformed bands and their excitations just coincide 
with the energies of the first T=3/2 states. In that case there is no isospin 
mixing between states of different isospin. Clearly, determination of the spins 
and parities of these states will answer this ambiguity.

\begin{figure}
\includegraphics[width=1.\textwidth]{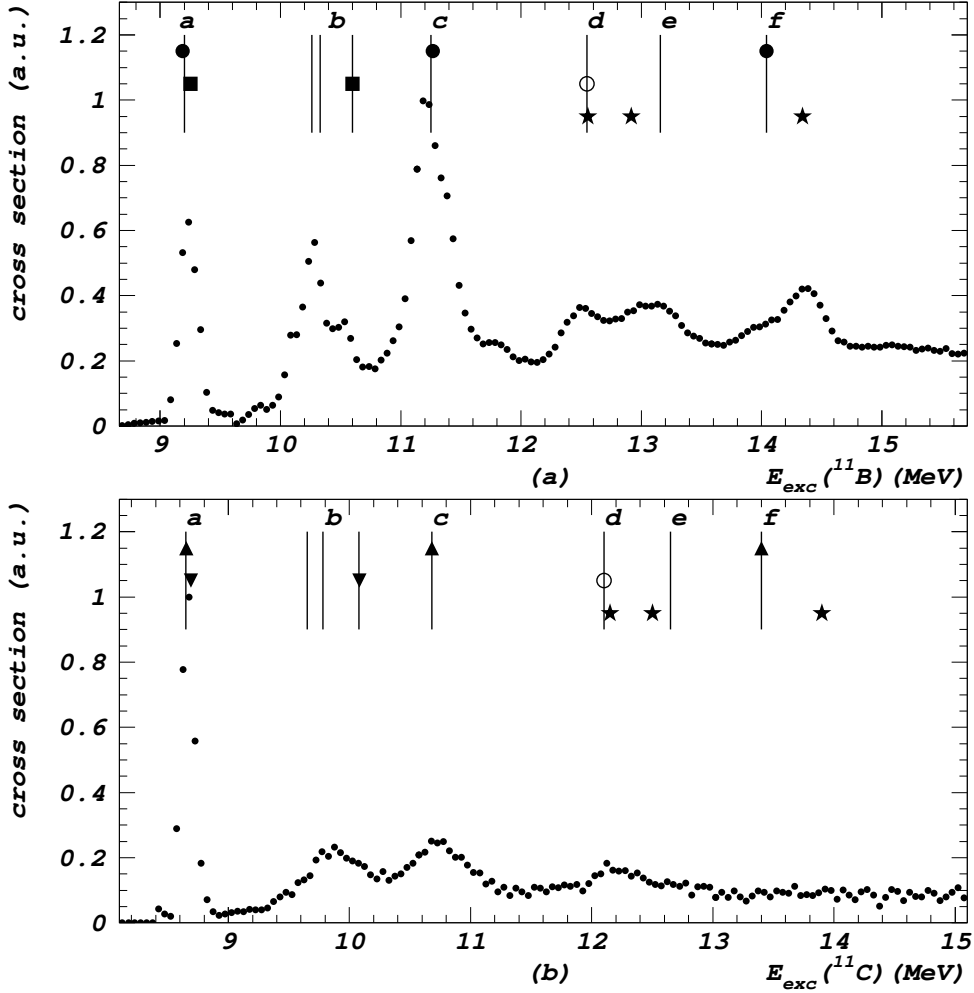}
\caption{\label{excs}
Comparison of the $^{11}$B and $^{11}$C excitation energy spectra from the
present measurements. Spectra are normalised for detection efficiency and
shifted so that the 7/2$^+$, 5/2$^+$ doublets are aligned. Lines mark 
positions of the states with the same value of the spin and parity 
populated in both nuclei. The positions of the bands members are marked 
with the same symbols as in Fig. \ref{ejj}. The excitations of the isobaric 
analogs of the lowest three $^{11}$Be states are marked with stars.  }
\end{figure}

An interesting feature of the present results (see Fig. \ref{excs}, which 
shows the efficiency corrected excitation energy spectra) is that we observe 
the same series of excited states at the lower excitations in both nuclei: 
unresolved doublet of 7/2$^+$, 5/2$^+$ states (marked 'a' in Fig. \ref{excs}), a 
3/2$^-$, 5/2$^-$, 7/2$^+$ triplet ('b' in Fig. \ref{excs}) which is $\sim$1.2 MeV 
above the doublet, 9/2$^+$ state ('c') at an excitation of 2 MeV higher then the 
doublet, then the proposed isobaric analogs of the $^{11}$Be ground state ('d') which 
are $\sim$3.4 MeV above the doublet, and then weak states, which are probably 
7/2$^+$ ('e') and 11/2$^+$ ('f'), and which are 3.9 and 
4.8 MeV above the doublet's excitation. All states observed in $^{11}$C appear
also as strong resonances in the $^{11}$B spectra. However, we have observed 
more states at higher excitations in $^{11}$B and also some weak states at 
lower excitations, which are mising in the $^{11}$C spectra. This is probably
due to the very different Q-value of the reactions used in the studies of the
$^{11}$B (Q=-2.461 MeV) and $^{11}$C (Q=-14.602 MeV) and the different 
kinematical conditions in the reactions. These strongly excited states observed 
in $\alpha$-decay of both the $^{11}$B and $^{11}$C should have the same 
structure. Observed strong $\alpha$+$^{7}$Li($\alpha$+$^{7}$Be) decay of 
these mirror states produced in the reactions involving transfer of two 
nucleons onto the 2$\alpha$+n cluster nucleus $^{9}$Be and known 
$\alpha$+$^{3}$H ($\alpha$+$^{3}$He) cluster structure of $^{7}$Li 
($^{7}$Be) suggest 2$\alpha$+$^{3}$H (2$\alpha$+$^{3}$He) three-centre 
cluster structure of the $^{11}$B ($^{11}$C) excited states. Support for 
existence of such structure in $^{11}$B and $^{11}$C can be found in the 
three-cluster Generator Coordinate Method calculations \cite{des}, simple 
cluster model calculations \cite{kabir} and three-cluster orthogonality 
condition model calculations \cite{furu} which provide a reasonable 
description of the $^{11}$B and $^{11}$C. This three-centre structure is 
also consistent with the results of the antisymmetrized molecular dynamics 
calculations of $^{11}$B \cite{kana95} and $^{11}$C \cite{kana97} which 
showed that even ground state and the lowest excited states possess deformed 
three-centre structure. 
The existing cluster models calculations have not examined the rotational
structures of the three-centre configurations and such calculations would
be extremely useful for the understanding of the $^{11}$B and $^{11}$C properties.

The Nilsson deformed single-particle level scheme indicates that
for oblate deformations the [202]5/2$^+$, [202]3/2$^+$ and
[200]1/2$^+$ orbits descend from the $sd$-shell. Excitations of a
proton (neutron) in $^{11}$B ($^{11}$C) to these orbits would
permit the formation of $K$=5/2$^+$ and 3/2$^+$ bands in the case
of the first two. This gives an indication that the observed
structures may be oblate in character. 

An additional neutron introduced into the 2$\alpha$+$^{3}$He system produces 
well known 3$\alpha$ structure of $^{12}$C which is then three-centre core 
for proposed molecular structures formed by addition of valence neutrons 
around it \cite{mil02,ita04,ita01}. In contrast, the 2$\alpha$+$^{3}$H
system in $^{11}$B is the most stable three-centre structure in boron
isotopes and addition of a neutron would result in molecular neutron orbital 
around that core. Indications for such structure in $^{12}$B have been
found in the $\alpha$+$^{8}$Li decay of excited states in the present
data \cite{cluster8} and some other studies of the $^{7}$Li+$^{9}$Be
$\longrightarrow$ 2$\alpha$+$^{8}$Li \cite{so12b,so12bfiz} and
$^{7}$Li($^{7}$Li,$^{8}$Li$\alpha$) \cite{fletcher} reactions.

\section{Summary}

Measurements of the $^{16}$O($^{9}$Be,$\alpha$$^{7}$Be)$^{14}$C and 
$^{7}$Li($^{9}$Be,$\alpha$$^{7}$Li)$^{5}$He reactions at E$_{beam}$=70
MeV provide evidence for $\alpha$+$^{7}$Be(gs) and $\alpha$+$^{7}$Li(gs)
decay of excited states in $^{11}$C and $^{11}$B. The
$^{11}$C excitation energy spectra provide evidence for resonances at 
8.65, 9.85, 10.7 and 12.1 MeV and indications for peaks at 12.6 and 13.4
MeV. This result is the first direct observation of $\alpha$-decay for
states above 9 MeV. The $^{11}$B excitation energy spectra 
show resonances at 9.2, 10.3, 10.55, 11.2, (11.4), 11.8, 12.5, (13.0), 13.1, 
(14.0), 14.35, (17.4) and (18.6) MeV. The observed $\alpha$+$^{7}$Li decay 
extends the excitation energy range in $^{11}$B for this decay channel. 
Given the nature of the reaction processes, 
two-nucleon transfer onto the 2$\alpha$+n cluster nucleus $^{9}$Be, and
the $\alpha$-decay of excited states at excitations where various decay
channels are possible, as well as known $\alpha$+t($^{3}$He) structure 
of $^{7}$Li($^{7}$Be), it is possible that these states are linked
with the three-centre 2$\alpha$+t($^{3}$He) cluster structure. This cluster 
structure appears to be more prominent in the positive-parity states, where 
two rotational bands corresponding to very deformed structure are suggested. 
The K=5/2$^{+}$ bands consist of 5/2$^{+}$, 7/2$^{+}$, 9/2$^{+}$ and 11/2$^{+}$ 
members and have rotational parameter $\hbar^{2}$/2I of 0.25 MeV. The rotational 
parameter of the K=3/2$^{+}$ bands with 3/2$^{+}$, 5/2$^{+}$, 7/2$^{+}$ and 
possible 9/2$^{+}$ members is 0.215 MeV. It is likely that these states are 
associated with oblate type structures similar to those found recently in 
the calculations of the rotational behaviour of $^{14}$C \cite{ita04}.

Excitations of some of the observed T=1/2 resonances coincide with the positions 
of T=3/2 states which are the isobaric analogue states of the lowest $^{11}$Be 
states, which would indicate mixed isospin. In this case, the states observed 
at excitations which correspond to the analogs of the $^{11}$Be ground state with 
J$^{\pi}$=1/2$^{+}$ may be the 1/2$^{+}$, T=1/2 states which mix with the analogues,
or alternatively the 9/2$^{+}$ members of the K=3/2$^{+}$ rotational bands. It is 
clear that the determination of the spins and parities of these states are imperative 
in order to understand structure of $^{11}$B and $^{11}$C excited states. Due to 
the large number of reaction amplitudes contributing to the reaction processes, 
resulting from the presence of nonzero spin nuclei in the entrance and exit channels, 
the angular distributions and angular correlations in the present measurements did 
not provide information on the spin and parity of the observed states. Additional 
measurements capable of determining such information are planned for the near future. 
The available theoretical calculations have not examined such three-centre systems,
where there are holes rather than particles being exchanged between $\alpha$-particles.
Additional calculations are important to further improve our understanding 
of the proposed structures.

{\bf Acknowledgments}

The authors would like to acknowledge the assistance of ANU personnel in
running the accelerator. NS is grateful to M. Milin for many useful discussions.
This work was carried out under a formal agreement
between the U.K. Engineering and Physical Sciences Research Council and the
Australian National University. PJL, BJG and KLJ would like to acknowledge the
EPSRC for financial support.

\end{document}